# Plasticity in irradiated FeCrAl nanopillars investigated using discrete dislocation dynamics

Yash Pachaury[1], George Warren[1], Janelle P. Wharry[1], Giacomo Po[2], Anter El-Azab[1]

[1]Purdue University, 205 Gates Road, West Lafayette, IN 47906, USA
[2]University of Miami, 1251 Memorial Drive, Coral Gables, FL 33146, USA

## Abstract

In this paper, we investigate plasticity in irradiated FeCrAl nanopillars using discrete dislocation dynamics simulations (DDD), with comparisons to transmission electron microscopic (TEM) *in situ* tensile tests of ion and neutron irradiated commercial FeCrAl alloy C35M. The effects of irradiation-induced defects, such as a/2<111> and a<100> type loops and composition fluctuations representative of phase separation in irradiated FeCrAl alloys, are investigated separately as well as superposed together in simulations. We explore the effects of defects on the stress-strain behavior, specifically yield strength and hardening response, of FeCrAl nanopillars. Our simulations confirm the widely accepted fact that irradiated alloys exhibit a stress-strain response with higher yield strength and hardening as compared to homogeneous alloys. However, our DDD calculations reveal an atypical superposition of the hardening contributions due to composition inhomogeneity and irradiation loops wherein hardening due composition inhomogeneity counteracts hardening due to irradiation loops at small scales. As a result, we observe that the yield strength in irradiated alloys, after taking into consideration the effects of both composition inhomogeneity and irradiation loops, is smaller than the yield strength of the alloys with only irradiation loops and is approximately same for the alloy with composition inhomogeneity alone. We identify this "destructive interference" in the superposition in our parallel TEM *in situ* tensile tests on unirradiated, ion irradiated, and neutron irradiated C35M FeCrAl alloy as well. This

Email Addresses: ypachaur@purdue.edu (Yash Pachaury), aelazab@purdue.edu (Anter El-Azab)

destructive interference in the hardening contributions contrasts with the widely utilized dispersed barrier hardening (DBH) models by the experimental community to model the hardening contributions due to different irradiation induced defects. Effects of the loading orientation on yield strength and hardening are investigated and the mechanisms for the hardening in irradiated FeCrAl alloys are also reported.

## 1. Introduction

Radiation-induced defects affect the plastic deformation behavior of crystalline materials. Plasticity in irradiated materials is a consequence of the collective dynamics of dislocations and their interactions with these defects. Defects interact locally with dislocations, typically hindering their motion, with collective effects which manifest in the form of hardening and loss of ductility (Churchman et al., 1957; Xiao, 2019; Xiao et al., 2020). Plastic behavior of irradiated materials becomes even more complex at the micron and submicron scales due to intrinsic and extrinsic size effects (Hosemann, 2018; Kiener et al., 2011; Wharry et al., 2019), in which pristine unirradiated micro-scale specimens appear harder than highly damage irradiated micro-scale specimens (Yano et al., 2017). These size effects can be explained by phenomena such as plastic instabilities arising from strain bursts (Ng and Ngan, 2008; Stefanos Papanikolaou et al., 2017; Qu et al., 2020; Wang et al., 2018, 2012; Zaiser et al., 2008), formation of defect-free channels (Edwards et al., 2005; Kacher et al., 2012), dislocation starvation (Greer and Nix, 2006; Jérusalem et al., 2012), and activation of dislocation sources (such as single-arm sources) (Cui et al., 2014; Parthasarathy et al., 2007), all of which introduce stochasticity, dependent on the sample preparation and conditions, in the mechanical behavior of materials (S. Papanikolaou et al., 2017).

This paper focuses on the submicron plasticity of ferritic FeCrAl alloys, which are candidate materials for nuclear power production applications (Field et al., 2018). Owing to their resistance to corrosion, high-temperature oxidation, stress-corrosion cracking, high-temperature structural degradation, and radiation-induced swelling, FeCrAl alloys are candidates for nuclear fission reactor cladding materials and some structural components (Edmondson et al., 2016; Field et al., 2018, 2015; Rebak et al., 2018, 2017). However, these alloys are susceptible to the formation of irradiation-induced dislocation loops and composition changes (phase separations which are precursors to, or are α' precipitations) (Aydogan et al., 2018; Briggs et al., 2017; Field et al., 2018, 2017, 2015; Massey et al., 2021). These irradiation-induced defects affect the mechanical behavior of the alloys which in turn degrades the structural integrity of the components made from these alloys, hence affecting their in-service operations and longevity. Therefore, it is vital to understand the mechanisms responsible for changing the mechanical behavior, specifically plasticity, in these alloys under the presence of such defects.

Plasticity in irradiated FeCrAl alloys arise due to dislocation-composition fluctuation (or α' precipitate) interactions, dislocation-irradiation loop interactions, and dislocation-dislocation interactions (Field et al., 2015). Different mechanisms superpose and give rise to the deformation behavior of the alloys observed experimentally. The superposed effect of different defects on the hardening response of the irradiated FeCrAl alloys has been typically rationalized using the dispersed barrier hardening model (DBH) and different superposition laws (Field et al., 2015; Mao et al., 2022; Massey et al., 2021). Field et al. (Field et al., 2015) used a linear and root-sum-square superposition law in conjunction with the DBH model and revealed that Cr-rich α' precipitates and

a/2<111> dislocation loops are weak barriers as compared to a<100> type dislocation loops. Their work revealed that a high density of even the weak α' precipitates introduce significant hardening in irradiated FeCrAl alloys. Moreover, an additive superposition in the hardening contributions due to different irradiation induced defects was shown in their work. Similar observations were reported by Massey et al. (Massey et al., 2021) and Mao et al. (Mao et al., 2022) using root mean squared superposition law. Owing to the similarities between irradiation induced microstructural evolution and mechanical behavior between Fe-Cr and Fe-Cr-Al alloys, it is believed that hardening models utilized for determining barrier strengths in FeCr alloys (Bergner et al., 2014; Tanigawa et al., 2009) can also be used for irradiated FeCrAl alloys.

Separating the effects of the different defects based on analytical hardening models involves an empirical fitting to the experimentally observed data which typically requires characterizing the defect densities. While the characterization of irradiation loops is typically carried out using transmission electron microscopy (TEM), characterization of the Cr-rich α' densities is typically performed using atom probe tomography (APT) by determining Cr-cluster iso-surfaces (Massey et al., 2021). The APT treatment ignores Cr-deficit α-Fe regions in predicting the overall hardening behavior of the alloys. In essence, prediction of hardening in irradiated FeCrAl alloys using analytical models does not consider the existence of the phase separated multicomponent solid solution, and rather treats it as a precipitation strengthened system. But in fact, hardening in irradiated alloys is due to the synergistic effects of the dislocation interactions with the Cr-rich and Cr-deficit regimes, along with their interactions with other defects. Since the composition fluctuation length scales in irradiated FeCrAl alloys are of the order of dislocation interaction distances (Pachaury et al., 2022b), discrete dislocation dynamics (DDD) can be utilized to probe plasticity in irradiated FeCrAl alloys by taking into consideration different irradiation-induced defects.

Additionally, small-scale mechanical testing of irradiated specimens has become a necessity due to limited irradiation volumes, limited space in reactors, and requirements for low radioactive waste (Hosemann, 2018; Lucas, 1990). But these small-scale specimens are subject to size effects, in which quantitative mechanical property measurements may not necessarily represent bulk-like mechanical properties, and in which the operative deformation mechanisms may be inconsistent with those in a bulk specimen. Therefore, it has become increasingly important to develop computational tools to explain the underlying mechanical behaviors of irradiated materials at small scales. With the recent algorithmic and computational advancements in the last two decades, DDD has emerged as a very powerful tool for probing plasticity at micron and sub-micron scales. It has been widely utilized to reveal size effects (Aitken et al., 2015; S. Papanikolaou et al., 2017), formation of defect-free channels (Cui et al., 2018a, 2018b; Nogaret et al., 2008), and plastic instabilities arising from bursts and dislocation avalanches at small scales in metals (Cui et al., 2021, 2017a, 2017b, 2016).

In this paper, we present a multiscale data-driven framework based on DDD for probing submicron plasticity in irradiated FeCrAl alloys. Taking cues from experimental observations, we model irradiation-induced defect loops and composition fluctuations which are representative of an irradiated FeCrAl alloy. We utilize a stochastic composition reconstruction scheme to generate composition fields in irradiated FeCrAl alloys. The generated composition fields are utilized to determine internal stresses in irradiated FeCrAl alloys arising from the composition fluctuations. Similarly, statistics acquired from experimental observations of the irradiation loops are also sampled into the DDD domain. We integrate the composition data through a composition-dependent mobility law and internal stresses, and irradiated loops into DDD to study the localized behavior of the dislocations at small scales. The effects of different irradiation-induced defects have been studied separately as well as superposed together to determine their relative influence on the plastic behavior of irradiated nanopillars. These results can partially explain TEM *in situ* tensile testing observations from neutron and ion irradiated FeCrAl alloy C35M also presented herein. In addition, the effects of orientation on the plasticity of irradiated FeCrAl alloys are also reported. The organization of the paper is as follows: a brief overview of the typical experimental observations regarding irradiation-induced defects in FeCrAl alloys is presented in Section 2 (2.1). Section 2 (2.2-2.4) also describes the outline of the data-driven multi-scale methodology employed in this study along with the computational and experimental details followed by results and discussions in Section 3 and conclusions in Section 4.

## 2. Methodology

### 2.1. Irradiation induced defects in FeCrAl alloys

Commercial FeCrAl alloy C35M is selected for this study, wherein the numerical digits in the alloy nomenclature indicate the nominal Cr and Al concentrations of 13 wt% and 5 wt%, respectively. The complete alloy concentrations are provided in Table 1. For the DDD model to capture microstructures representative of irradiation-induced defects, we utilize earlier experimental results of Massey et al. (Massey et al., 2021) on neutron-irradiated C35M together with our ion irradiated experimental results on the same process heat of C35M. The ion irradiation is conducted using 4.4 MeV $Fe^{2+}$ ions at 370 ± 5ºC using a 3 MV NEC Pelletron accelerator at the Michigan Ion Beam Laboratory, to an accumulated irradiation dose of 7 displacements per atom (dpa) at a depth of 400-600 nm from the damage peak. Both neutron and ion irradiations are selected because of the tendency for neutrons to create both loops and phase separation, while ion irradiation is known to nucleate loops while limiting the extent of Cr-rich α' phase separation (Swenson and Wharry, 2020, 2017). Table 2 shows the loop statistics and α' phase sizes (radius) and number densities for both the neutron-irradiated and ion- irradiated C35M. Figure 1 shows dislocation loops and elemental concentration as revealed by down-zone axis bright field scanning TEM (STEM) (Parish et al., 2015) and STEM energy dispersive X-ray (EDX) spectroscopy chemical mapping (Briggs, 2016), respectively, in ion-irradiated C35M. Irradiation induced defects as observed from STEM and APT in neutron irradiated C35M can be found in the earlier work by Massey et al. (Massey et al., 2021). Taking cues from statistical datasets for irradiation loops densities and composition fluctuations, we include these effects in DDD. We now move towards the DDD computational details followed by incorporating irradiation defect statistics into DDD and the simulation details for studying submicron plasticity in irradiated FeCrAl alloys.

Table 1. Alloy concentration (wt%); all other elements (Zr, B, Hf, V, W, Ce, Co, Cu, La, Mn, Ni) measured at or below <0.01.

| Alloy | Fe | Cr | Al | Y | Mo | Si | C | S | O | N | P |
|---|---|---|---|---|---|---|---|---|---|---|---|
| C35M | 79.43 | 13.06 | 5.31 | 0.053 | 2 | 0.13 | 0.001 | <0.0003 | 0.0012 | 0.0003 | 0.007 |

Table 2. Irradiation conditions and irradiation defect statistics for irradiated C35M alloy

| Irradiating Particle | Dose [dpa] | $T_{irr}$ [°C] | Dislocation loop number density ($\times 10^{21}$) [m$^{-3}$] | | Loop diameter [nm] | | α' phase number density ($\times 10^{24}$) [m$^{-3}$] | α' phase size, R [nm] | Source |
|---|---|---|---|---|---|---|---|---|---|
| | | | a<100> | a/2<111> | a<100> | a/2<111> | | | |
| Neutron | 1.8 | 214 | 6.34 ± 0.78 | 20.2 ± 3.17 | 13.4 ± 3.7 | 15.6 ± 6.8 | 1.9 ± 0.4 | 1.4 ± 0.3 | (Massey et al., 2021) |
| Neutron | 1.8 | 357 | 0.88 ± 0.28 | 2.47 ± 0.36 | 21.3 ± 11.1 | 20.4 ± 13.6 | 1.8 ± 0.3 | 2.2 ± 0.7 | (Massey et al., 2021) |
| 4.4 MeV Fe$^{2+}$ | 7 | 370 | 1.1 ± 0.6 | | 16.6 ± 4.7 | | none | none | This study |

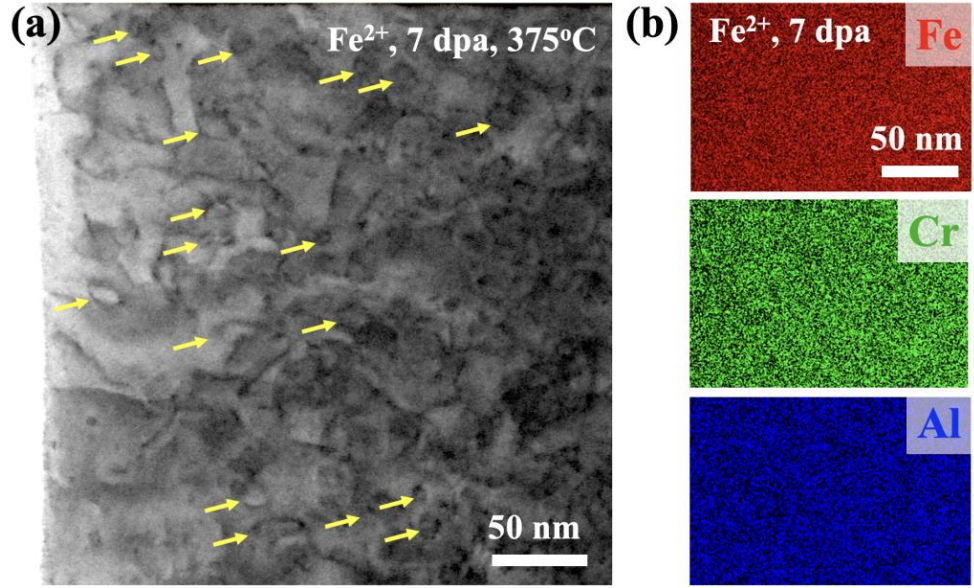

Figure 1: Bright field down-zone STEM micrographs of (a) $Fe^{2+}$ irradiated C35M, showing irradiation-induced nucleation of dislocation loops (arrowed) and (b) STEM EDX maps of $Fe^{2+}$ irradiated C35M showing compositional homogeneity of major alloying elements Fe, Cr, and Al, and lack of irradiation-induced α– α' phase separation.

## 2.2. Discrete dislocation dynamics for investigating submicron plasticity in FeCrAl alloys

We utilize Mechanics of Defect Evolution Library (MoDELib) for setting up the simulations of collective dynamics of dislocations in nanopillars (Po et al., 2014). In our modeling framework, closed dislocation loops are represented as a set of connected straight dislocation segments having a non-singular stress field solution of Cai et al. (Cai et al., 2006) with a core-size of 2b, where b represents the magnitude of the Burger's vector at the nominal composition of C35M. All simulations are carried out at room temperature. Molecular Statics (MS) calculations are utilized to determine the magnitude of the Burgers vector (b=0.247 nm) using the interatomic potential of Liu et al. (Liu et al., 2019) whereas the available experimental dataset from Field and coworkers (Field et al., 2018) is used to determine the elastic properties (shear modulus, G=77.88 GPa and Poisson's ratio, ν=0.27) at room temperature. To account for free surfaces in the nanopillars, an elastic boundary value problem (BVP), based on superposition principle of linear elasticity, is solved using finite element method (FEM) to determine image stresses. More details about the elastic BVP can be found in (Giessen and Needleman, 1995; Po et al., 2014; Weygand et al., 2002). Virtual segments outside of the simulation domain are utilized to close dislocation segments accumulating at the boundaries. This minimizes errors in calculating tractions at the boundaries of the simulation domain. Correspondingly, applied stress, dislocation-dislocation interactions, and image stresses act as the driving force which govern the evolution of the dislocations in the nanopillars. Other driving forces, such as the stress fields due to irradiation loops and composition fluctuations, also exist which also affect the glide dislocation motion in the domain. Incorporation of irradiation induced defects (loops and composition fluctuations) will be discussed in detail in the next subsection. Adaptive dislocation remeshing, junction formation, and cross-slip for the dislocation segments are checked at every iteration for possible reactions of the dislocation networks. The probabilistic cross-slip law of Chaussidon et al. (Chaussidon et al., 2008) for BCC metals is utilized in the current study. Now, we discuss the incorporation of irradiation-induced defects in our DDD framework.

## 2.3. Incorporating irradiation-induced defects in DDD

As discussed before, plasticity in irradiated FeCrAl alloys is a manifestation of the interactions of glide dislocations with fluctuating composition fields and irradiation loops. To effectively capture these interactions using DDD, it is essential that the composition fields and the irradiation loops be effectively sampled in the simulation domain such that they are statistically representative of typical observations from experiments. We utilize a stochastic reconstruction scheme for effectively capturing the experimentally observed probability distribution functions (PDFs) and composition covariances in the DDD simulation domain (Pachaury et al., 2022b). The stochastic composition reconstruction scheme is based on determining PDFs and covariances (both direct and cross-covariances), henceforth called experimental covariances, from experimental 2D EDX maps of irradiated FeCrAl alloys. The experimental covariances are fitted to parametric covariance functions which are representative of the second-order statistical information in the composition maps. Consequently, a spectral method is utilized for sampling the covariance information and reconstructing the composition data in 3D DDD domain. Utilization of 2D composition data to reconstruct 3D data assumes isotropy in the $3^{rd}$ direction in 2D composition maps. Interested readers are referred to Pachaury et al. (Pachaury et al., 2022b) for more details about the stochastic reconstruction scheme. In the current work, the Cr-composition covariance relaxation length scale, $\lambda$, was observed to be approximately 5 nm from experimental composition maps. $\lambda$ refers to the length scale beyond which the covariance relaxes to or oscillates about 0 (Pachaury et al., 2022b). For effectively modeling such small relaxation length scales, the dislocation discretization length scale must be smaller than $\lambda$. However, it was determined that such small dislocation discretization length scales led to very slow computational times for effectively capturing short-range discrete glide events (junction formation and cross-slip). Therefore, we have artificially increased $\lambda$ to approximately 50 nm which is 10 times larger than the experimentally observed $\lambda$. To quantify the changes that may arise while using artificial relaxation length scales, we have also studied the effects of $\lambda$ on plasticity in irradiated FeCrAl alloys. Composition profiles with $\lambda$ of 10nm, 30 nm and 75 nm are also studied which represent 2, 6 and 15 times, respectively, larger covariance relaxation length scales as compared to the experiments. To effectively capture the effects of composition fluctuations, the dislocation segments are discretized such that their length always remained between 5 nm and 15 nm throughout the course of all simulations. Each dislocation segment is discretized with 1 quadrature point for every 5 Burgers vector units of length. Composition and stresses are sampled at each quadrature point which in turn are utilized to determine the velocity at the quadrature points using a composition-dependent mobility law (Kumagai et al., 2021). These velocities are then used to determine the velocities of the dislocation nodes using a numerical Galerkin scheme. More details regarding the determination of velocities of the dislocation nodes using quadrature point velocities can be found in (Marian et al., 2020; Po et al., 2014). Now, we turn our attention towards composition dependent mobility law and the determination of the internal stresses due to fluctuating lattice parameters in FeCrAl alloys.

Kumagai et al. (Kumagai et al., 2021) carried out molecular dynamics (MD) calculations for determining the composition, stress, temperature, and line-direction dependence of the dislocation velocities in FeCrAl alloys. Parametric forms for screw, edge, and mixed dislocations were proposed where the parameters were found to be dependent on temperature, line-direction, and composition. The parametric form is implemented in the current framework to specify the mobility law for the dislocations. The fluctuating composition fields also give rise to fluctuating lattice parameters throughout the domain. This gives rise to internal stresses, henceforth called coherency stresses, which depend on the composition fields within the domain. The coherency stresses arise to compensate for the incompatibility introduced by stacking of crystallite chunks of different lattice parameters. The coherency stresses are determined by solving an elastic boundary value problem (BVP) such that the net equilibrium of the crystal is attained under internal stresses due to the fluctuating lattice parameter. Consequently, the BVP problem is solved by considering the domain to be an RVE of the bulk crystal. The BVP to solve for the internal coherency stresses, $\boldsymbol{\sigma_c}$ is:

$$\nabla \cdot \boldsymbol{\sigma}_C = 0, \tag{1}$$

such that

$$\mathbf{u} = \bar{\mathbf{u}} \text{ on } \partial\Omega_u \tag{2}$$

where $\boldsymbol{\sigma}_C = \mathbf{C} : \boldsymbol{\varepsilon}^e(\mathbf{x})$. Here, $\boldsymbol{\varepsilon}^e(\mathbf{x})$ is the elastic strain tensor. The total strain tensor for the problem can be written as $\boldsymbol{\varepsilon}(\mathbf{x}) = \boldsymbol{\varepsilon}^e(\mathbf{x}) + \boldsymbol{\varepsilon}^*(\mathbf{x})$, where $\boldsymbol{\varepsilon}^*(\mathbf{x})$ is the eigenstrain due to the fluctuating lattice parameter and is given by $\frac{1}{3}\frac{\Delta V}{V_{nom}}\delta_{ij}\, \mathbf{e_i} \otimes \mathbf{e_j}$. Here, $\Delta V$ represents the volume change of the lattice with respect to the nominal volume, $V_{nom}$. $V_{nom}$ corresponds to the crystal volume of the nominal alloy concentration of C35M. In Eq. (2), $\partial\Omega_u$ represents the boundary of the simulation domain. The implementation of the BVP is carried out in MOOSE (Permann et al., 2020) using grid spacing defined by the reconstructed composition maps. It is worthwhile to note that the composition fields as well as the coherency stresses are taken to be constant within each voxel. Now, we turn our attention towards incorporating irradiation induced loops in DDD.

Both a/2<111> and a<100> type irradiation loops are observed during irradiation of FeCrAl alloys. Taking cues from Table 2, both a/2<111> and a<100> are sampled in the DDD domain representative of the FeCrAl nanopillars such that the number densities of a/2<111> and a<100> are $2.47 \times 10^{21}$ m$^{-3}$ and $8.8 \times 10^{20}$ m$^{-3}$ respectively. Earlier work by Briggs et al. (Briggs, 2016) did not find any correlation between composition fluctuations and the nucleation of irradiation loops. Therefore, both types of loops are generated randomly from a Gaussian distribution with mean diameter of 20.4 nm and a standard deviation of 13.6 nm such that these resemble to be generated after neutron irradiation of FeCrAl alloys to an accumulated irradiation dose of 1.8 dpa at 357 °C. a/2<111> type irradiation loops are generated as hexagonal loops whereas a<100> loops are generated as square loops based on earlier works on characterization of irradiation loops in α-Fe (Cui et al., 2017b; Eyre and Bullough, 1965; Fitzgerald and Yao, 2009). Figure 2 shows a schematic representation for the multiscale data-driven framework adopted in this study where 3D composition fields, their coherency stresses, and a parametric form of the dislocation velocities are coupled to DDD where irradiation induced loops were sampled randomly in the domain.

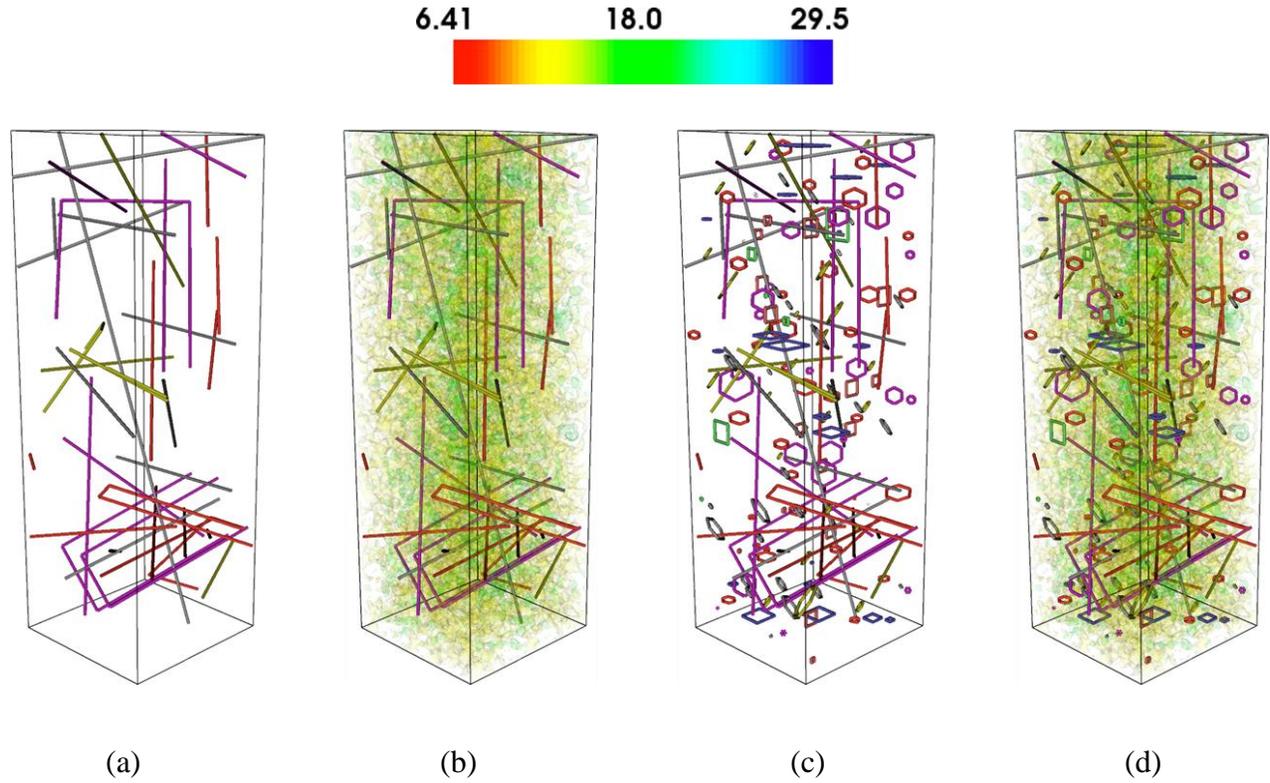

(a)          (b)          (c)          (d)

Figure 2: Schematic for different defects (glide dislocation and irradiation induced defects) utilized in DDD. (a) glide dislocations (b) composition fields showing the fluctuations in the Cr-composition superposed with glide dislocations, (c) irradiation induced $a/2<111>$ (shown as hexagons) and $a<100>$ (shown as square) loops superposed with glide dislocations, and (d) all defects (glide dislocations, irradiation induced composition changes and loops) superposed together and representative of irradiated FeCrAl pillars. Color bar corresponds to the composition fluctuations in Cr wt.%.

## 2.4. Other simulation and experimental details

While we have already discussed many aspects of the DDD problem formulations and the corresponding inputs in Sections 2.2 and 2.3, we present additional details regarding the DDD simulations in this section. A simulation domain of 250 nm × 250 nm × 250 nm is utilized. The nanopillars are loaded under strain loading control at a constant strain rate of 100 s$^{-1}$ along the z direction. The bottom surface of the nanopillars is fixed by restricting the degrees of freedom of the nodes at the bottom surface whereas the nodes at the top surface are displaced. Only glide in $\{110\}\langle 1\bar{1}1\rangle$ slip systems is considered in the present study. Three different classes of problems are investigated to reveal nanoscale plasticity in irradiated FeCrAl alloys: 1. separate vs combined effects to reveal relative influence of the irradiation induced defects on yield strength (hardening), 2. effects of the composition covariance relaxation length scales on stress-strain response and yield strength, and 3. effects of the loading orientation on yield strength and plasticity. Corresponding to each class of problem, 3 different simulations are carried out with slightly varying starting glide densities of $1.8\times10^{14}$ m$^{-2}$, $2\times10^{14}$ m$^{-2}$, and $2.2\times10^{14}$ m$^{-2}$ to bring in the stochastic nature of the deformation behavior at small scales. Initial dislocation microstructure consists of straight dislocations and single arm sources comprising of 75% and 25%, respectively of the total glide dislocation density. The crystal is oriented along [100]×[010]×[001] for all the simulation conditions studied except where we analyze orientation-dependent submicron plasticity. For analyzing orientation-dependent plasticity in the nanopillars, we have used two other crystal orientations along [$\bar{1}$11]×[1$\bar{1}$2]×[110] and [$\bar{1}$10]×[$\bar{1}\bar{1}$2]×[111] crystal directions. It is worthwhile to note that the same composition fluctuations (with λ of 50 nm) are utilized for the different orientations of the crystal. Prior to loading, the dislocations in each simulation are relaxed under the internal stress fields (due to dislocation-dislocation, dislocation-composition fluctuations, and dislocation-irradiation loop interactions) to an extent that no further motion of the dislocations was observed with increase in simulation time.

We also conduct *in situ* tensile tests of as-received, neutron irradiated, and ion irradiated C35M specimens to probe nanoscale plastic deformation from an experimental perspective. It is to be noted that only ⟨110⟩ orientation loading is utilized to study the deformation of as-received, neutron and ion irradiated FeCrAl alloys. First, grains with ⟨110⟩ orientations are identified by scanning electron microscopy (SEM) with electron backscatter diffraction (EBSD). Subsequently, focused ion beam (FIB) milling on the selected grains is used to extract lamellae from the bulk specimens, then shape them into tensile dogbone specimens on push-to-pull TEM *in situ* tensile device following the process described in (Warren et al., 2020). Depth-sensing TEM *in situ* tensile testing was conducted using a Bruker (formerly Hysitron) PI-95 Picoindenter TEM *in situ* holder with a 20 µm diamond flat punch at a constant strain rate of 6 x $10^{-4}$ $s^{-1}$. The results obtained from DDD are not benchmarked against these tensile tests but will rather be used to help explain the results of these tensile experiments qualitatively. Now we turn our attention towards insights gained from DDD simulations for plasticity in FeCrAl nanopillars.

## 3. Results and discussions

In this section, we present the results for the DDD simulations for the 3 classes of problems studied in this paper. H, I, HL, and IL are used as short-hand notations to represent results for homogeneous, inhomogeneous (composition fluctuations effects only), homogeneous with irradiation loops (irradiation loops effects only), and inhomogeneous with irradiation loops (combined effects with composition fluctuations and irradiation loops), respectively. Orientations and composition covariance relaxation length scales ($\lambda$) are indicated wherever necessary.

Otherwise, all simulation results correspond to the crystal orientation along [100]×[010]×[001] and composition covariance relaxation length scale (λ) of 50 nm.

Figure 3 shows the stress-strain behavior of the nanopillars across all the runs showing the effects of different defects separately as well as superposed together. Respective runs for different cases correspond to the same initial dislocation microstructure. Irradiation induced dislocation microstructure are also same for HL and IL runs. Different runs exhibit different hardening behavior for the alloys. This observation is in tandem with stochasticity in mechanical testing of materials at small scales. Hardening in the nanopillars without irradiation loops (see Run 3 in Figure 3(a)) is observed to be due to intermittent dislocation starvation events occurring after dislocation sweeping the entire slip plane and getting deposited at the boundary and subsequent activation of other dislocations at higher stresses. Irradiation induced defects act as barriers to the dislocation motion due to which hardening in irradiated alloys is a manifestation of interaction of dislocation with irradiation induced defects in addition to the dislocation starvation and activation. A characteristic dislocation starvation event is observed during the deformation of inhomogeneous alloy (I) as shown by the yellow shaded region in Figure 3(b). The corresponding event happening is also shown in Figure 4. Two arms of dislocation sources which are contributing to plasticity in the nanopillars (Figure 4 (a)) annihilate due to cross-slip acquiring a small closed configuration (Figure 4 (b)). The corresponding effect is manifested in the stress-strain curve as shown by an upward triangle in Figure 3 (b). As a result, no dislocations move and contribute to the plastic distortion in the crystal as observed by linear increase in the stress with strain. As the stresses in the nanopillar increase, other dislocation sources (grey source and magenta loop in Figure 4 (c)) are activated. As a result, the nanopillar starts deforming plastically and the corresponding effect is observed in the stress-strain behavior as shown after the triangle pointing downwards in Figure 3 (b).

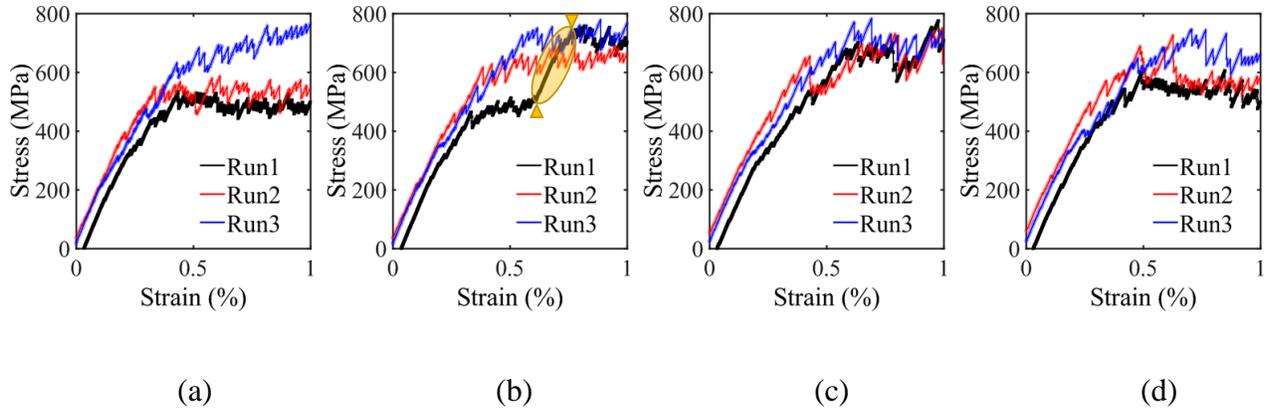

(a) (b) (c) (d)

Figure 3: Stress-strain curves obtained from DDD simulations of nanopillars for: (a) homogeneous alloy (H), (b) inhomogeneous alloy (I), (c) homogeneous alloy with irradiation loops only (HL), and (d) inhomogeneous alloy with both composition fluctuations and irradiation loops (IL). In (b) yellow region corresponds to dislocation annihilation and starvation event (shown by a triangle pointing upwards) followed by hardening and activation of other dislocations at higher stresses (shown by a triangle pointing downwards).

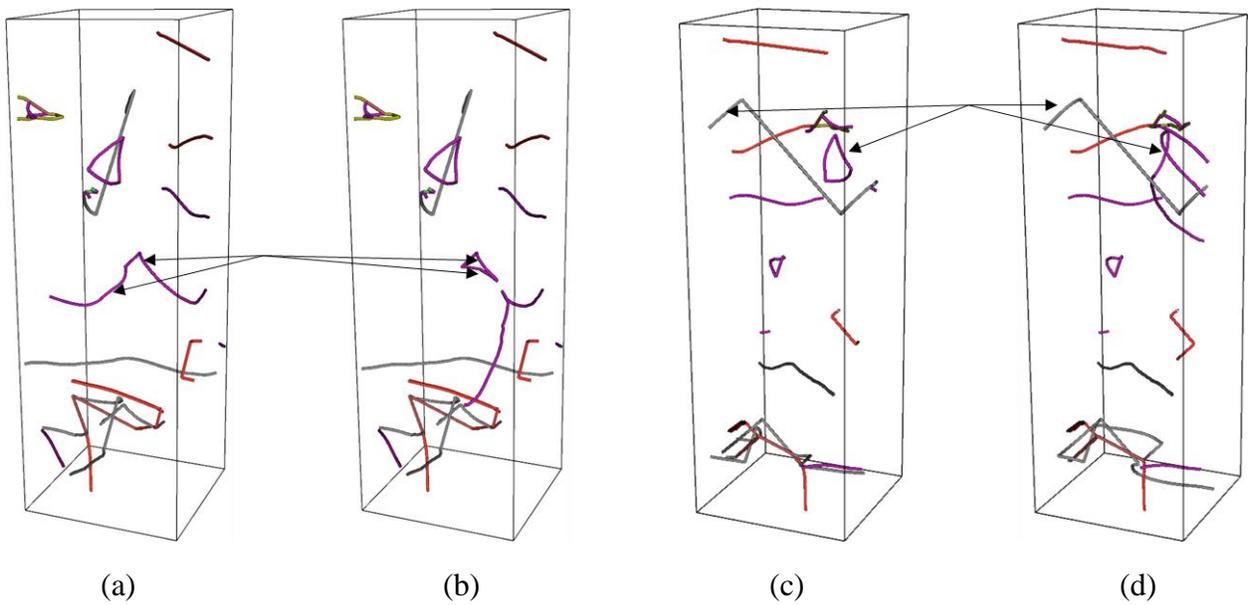

(a) (b) (c) (d)

Figure 4: Dislocation starvation event happening in inhomogeneous alloy (I), (a) dislocations marked by arrow indicate active dislocation sources, (b) Bottom arm cross-slips and annihilates with the top arm (corresponding arrows from (a) are shown on (b)). Previously active arms acquire a small closed configuration and become inactive due to their small size., (c) Activity of dislocations at higher applied stresses. Arrows in the figure show the dislocations which become active at higher stress, and (d) plastic activity after activation of the other dislocations (corresponding arrows from (c) are shown on (d)). The right two figures ((c) and (d)) are rotated by approximately 90 ° about the vertical direction as compared to the first two ((a) and (b)). (a) and (b) correspond to the region pointed by upward triangle in Figure 3(b) whereas (c) and (d) correspond to the region pointed by downward triangle in Figure 3(b) in the yellow shaded region.

Figure 5 shows the separate vs combined effect through an averaged response taken across all the runs for FeCrAl nanopillars. Averaged stress-strain behavior is shown is Figure 5(a) whereas 0.2% yield strength is shown in Figure 5(b). It is worthwhile to note that the error bars in Figure 5(b) represent the minimum and maximum value of the yield strength observed from the simulations. Hardening is evident through an increase in the yield strength in nanopillars with irradiation induced defects as compared to the homogeneous alloy. However, an interesting observation is the reduction in the yield strength for the simulations carried out taking into consideration the effects of both composition fluctuations and irradiation loops as compared to the ones carried out by considering only the effects due to irradiation loops. Moreover, the average yield strength under the presence of both irradiation loops and composition fluctuations is observed to be similar to the yield strength under the presence of composition fluctuations except for the variability. Irradiation loops act as strong barriers to the motion of the dislocations as compared to the composition fluctuations. They can pin the dislocations gliding along their path and can also decorate the dislocations which require more stress to activate the dislocations. However, composition fluctuations generate local coherency stresses which provide an additional driving force to the gliding dislocations and a/2<111> type irradiation loops. As a result, the interactions between the glide dislocations and a/2 <111> type loops changes for inhomogeneous alloy (IL) as compared to the homogeneous alloy (HL) with irradiation loops. Figure 6 shows an interaction of an a/2[1$\bar{1}$1] type straight dislocation with an a/2[$\bar{1}$11] type irradiation loop for homogeneous alloy (HL) (Figure 6 (b)) and inhomogeneous alloy (IL) with irradiation loops (Figure 6(c)) after relaxation under their own internal stress fields. The corresponding starting configuration is shown in Figure 6 (a). The two rows show two different angles from which the interaction could be understood. For the HL alloy configuration, the two dislocations relax but no topological changes are observed whereas the two dislocation react and a part of the loop forms an a[001] type sessile junction for the IL alloy configuration. Additionally, Figure 6 also illustrates the changes in the orientation of an a/2[$\bar{1}$1$\bar{1}$] type irradiation loop (shown in magenta in Figure 6 (a) bottom row) due to the presence of the composition fluctuations. The rotation of the loop is larger under the presence of the composition fluctuation as compared to the simulation where the composition fluctuations are absent. Therefore, changes in the configurations of the a/2⟨111⟩ type irradiation loops are believed to be the reason for the observation of "destructive interference" between hardening due to irradiation loops and composition fluctuations. It is worthwhile to note that this observation contrasts with the typical experimental observations (using DBH models) on characterizing the relative impacts of different irradiation induced defects on the overall hardening of irradiated FeCrAl alloys where both irradiation loops and composition fluctuations contribute positively to the overall hardening of the alloy (Field et al., 2015; Mao et al., 2022; Massey et al., 2021). In other words, a "constructive interference" is observed in the experimental characterization of the hardening of irradiated FeCrAl alloys. A detailed analysis comprising of experiments and simulations for determining the relative influence of different irradiation induced defects is beyond the scope of the current work.

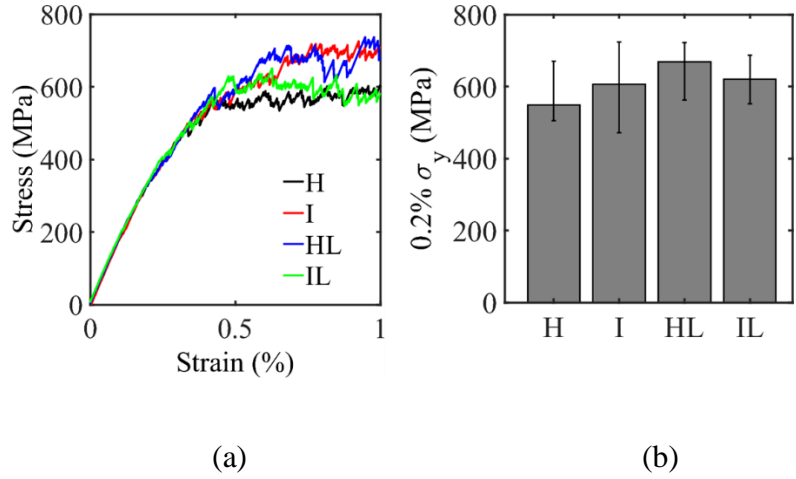

(a)  (b)

Figure 5: Separate vs combined (a) averaged stress-strain response of FeCrAl alloys, and (b) 0.2% yield strength as determined from DDD.

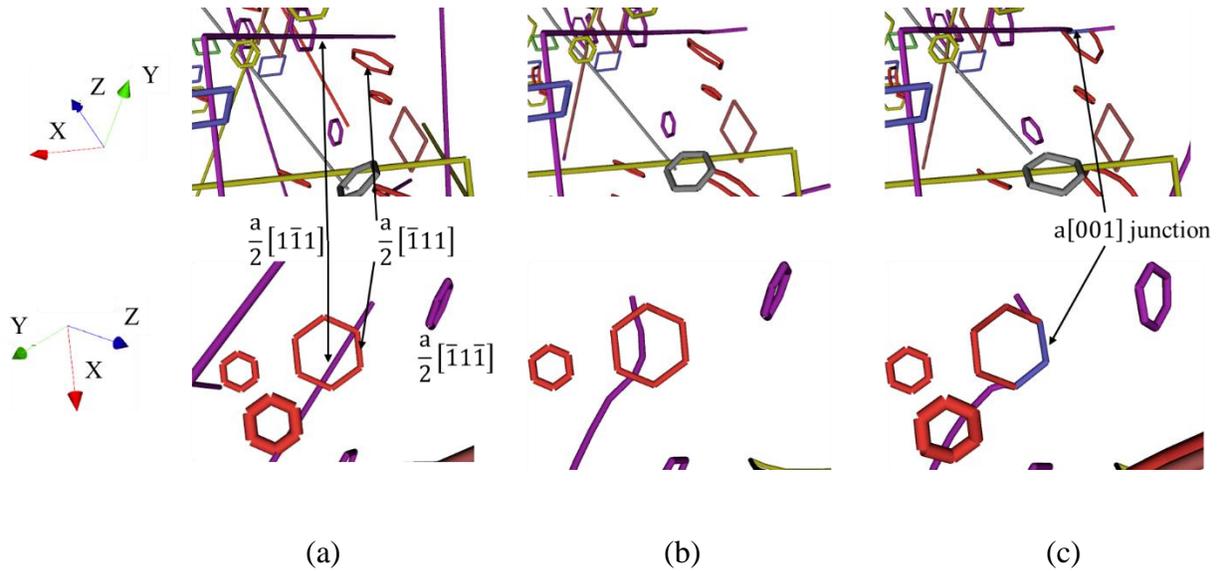

(a)  (b)  (c)

Figure 6: Interaction of a glide dislocation with irradiation loops and the difference observed in dislocation network topology due to the composition fields. (a) Initial dislocation configuration,

(b) dislocation configuration after relaxation for homogeneous alloy with loops (HL), and (c) dislocation configuration after relaxation for the inhomogeneous alloy with loops (IL). The glide dislocation forms a junction of type $a[100]$ with the irradiation loop for IL alloy configuration whereas no junction formation is observed for the HL alloy configuration. Additionally, significant rotation of $a/2[\bar{1}1\bar{1}]$ type loop (shown in purple and labelled in the bottom row) can be observed due to internal stresses of the composition fields. The top row and bottom row show the same dislocations interacting but in two different viewing orientations to clearly depict the topological changes in the domain.

Nevertheless, results for the experimental TEM *in situ* tensile testing as shown in Figure 7 can be explained by this modeled synergistic effect of loops and composition inhomogeneities on overall hardening. Quantitatively, the specimens are size effected, resulting in artificially elevated yield and flow stresses than their bulk counterparts. The size effect is most pronounced in the unirradiated specimen due to its lack of dislocation sources, resulting in the appearance of irradiation softening, which is typical of irradiated metals in small-scale mechanical testing (Hosemann, 2018; Kiener et al., 2011; Wharry et al., 2019). With regard to the irradiated specimens, the ion irradiated specimen exhibits a slightly higher yield strength than the neutron irradiated specimen. This result is consistent with the DDD prediction that loops and composition inhomogeneities superpose in an anti-correlative manner (destructive interference) to cause apparent softening (e.g., in the neutron irradiated case), compared to loops alone in a homogenous alloy (e.g., ion irradiated case). Nevertheless, it is worth considering these individual results in the context of the stochastic nature of micro-mechanical tests, which would suggest that hundreds of duplicate specimens may be necessary to obtain statistical certainty in results (Malyar et al., 2018).

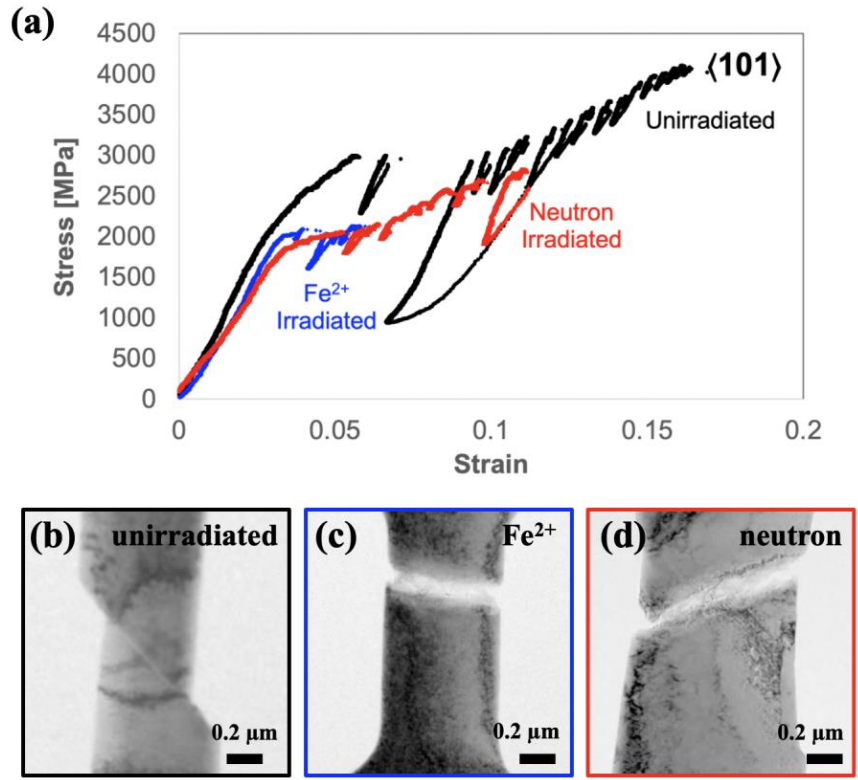

Figure 7: Comparison of push-to-pull TEM in situ tensile testing of unirradiated, ion irradiated, and neutron irradiated C35M. (a) Quantitative stress-strain curves reveal that the unirradiated specimen is size effected, while the ion irradiated specimen exhibits slightly higher yield stress than the neutron irradiated specimen. TEM still-frame images taken from the end of each in situ tensile test are shown for (b) unirradiated, (c) ion irradiated, and (d) neutron irradiated specimens.

To further confirm this destructive interference between superposition of loops and composition inhomogeneities, new sets of simulations were carried out for the HL and IL alloy configurations by making the a/2<111> type irradiation loops immobile. It is noteworthy that the irradiation microstructures were different from the simulations presented in Figure 5 but the glide dislocation microstructure was kept the same. Figure 8 shows the averaged stress-strain response (Figure 8 (a)) and 0.2% yield strength (Figure 8 (b)) for the simulations. It is evident that IL simulations have the highest yield strength due to composition fluctuations not changing the morphology and orientations of a/2<111> type irradiation loops as a result of which the interaction of the irradiated microstructure is similar for both HL and IL type simulations.

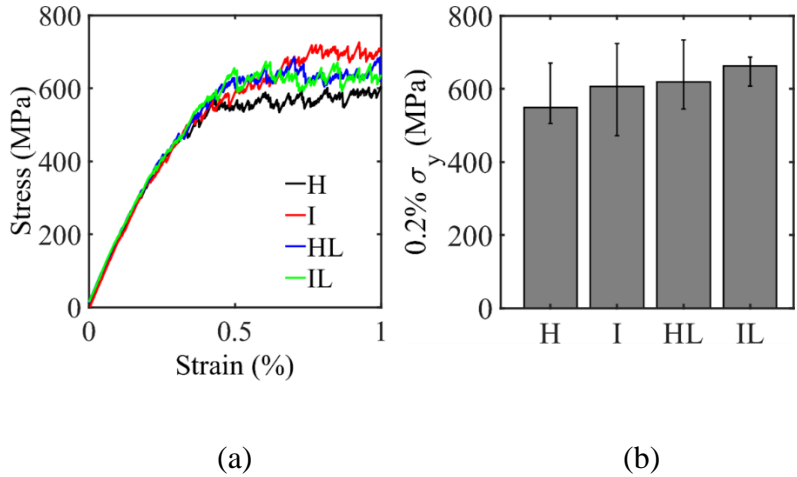

(a)                            (b)

Figure 8: Separate vs combined (a) averaged stress-strain response of FeCrAl alloys, and (b) 0.2% yield strength as determined from DDD considering a/2<111> type irradiation loops to be immobile.

Figure 9 shows the dislocation densities at 1% strain in the four alloy configurations tested. Figure 9 (a) shows the initial dislocation microstructure for homogeneous (H) and inhomogeneous (I) alloy configurations for which the dislocation microstructures at 1% strain are shown in Figure 9 (b) and (c) respectively. Figure 9 (d) shows the initial dislocation microstructure for homogeneous (HL) and inhomogeneous (IL) alloy configurations with irradiation loops for which the dislocation microstructures at 1% strain are shown in Figure 9 (e) and (f) respectively. It is evident that the evolution of the dislocation density is different across all the alloy configurations tested. Glide dislocations interact with irradiation loops where they get pinned. In other instances, the irradiation loops are destroyed (or consumed) by the gliding dislocations by formation of junctions. Composition fields and irradiation microstructure also change the local stress fields around the glide dislocation which alters the cross-slip activity in the domain. Moreover, interactions of the glide dislocation with irradiation loops result in the generation of additional pinning points from where dislocation sources may nucleate into the nanopillars. These mechanisms are expected to alter dislocation densities in irradiated alloy as compared to homogeneous alloy. Now, we turn our attention towards investigating the effects of the composition relaxation length scales (λ) on the mechanical behavior of the irradiated nanopillars (IL).

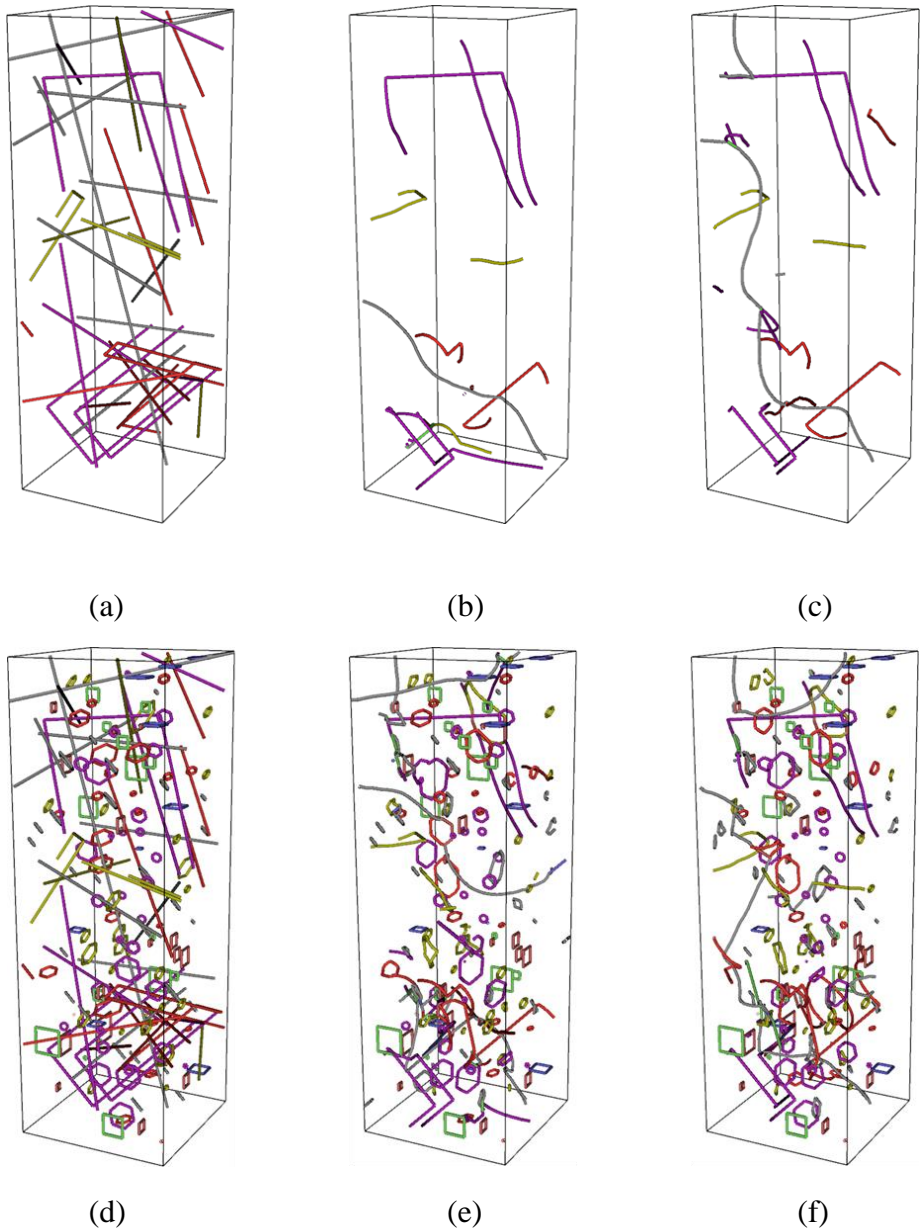

(a)            (b)            (c)

(d)            (e)            (f)

Figure 9: Dislocation density in different alloys. (a) Starting configuration for homogeneous (H) and inhomogeneous (I) alloys, (b) dislocation density in the homogeneous (H) alloy after 1% strain, (c) dislocation density in inhomogeneous (I) alloy after 1% strain, (d) starting configuration for homogeneous (HL) and inhomogeneous (IL) alloys with irradiation loops, (e) dislocation density in homogeneous (HL) alloy with irradiation loops after 1% strain, and (f) dislocation density in inhomogeneous (IL) alloy with irradiation loops after 1% strain.

As outlined earlier, we have taken the effects of irradiation-induced composition fluctuations into consideration in our DDD model by artificially increasing composition covariance relaxation length scales ($\lambda$). We have performed a set of simulations targeted towards investigating the effects of $\lambda$ on the mechanical behavior of the irradiated nanopillars. Effects of $\lambda$ are investigated under the presence of irradiation loops as well as in their absence in the simulation domain. Figure 10 shows the iso-surface Cr composition maps at 18 wt.% for the 4 different covariance relaxation length scales tested in this work. As is evident from Figure 10, the extent of the composition clustering increases with increase in $\lambda$. However, it was determined that $\lambda$ does not affect the stress-strain behavior of the irradiated alloy significantly. Figure 11 shows the stress-strain behavior of the irradiated alloy (IL) as a function of $\lambda$. Slight differences in the evolution of the stresses with strains are observed but the observation is inconclusive for the effects of composition covariance relaxation length scales on the stress-strain response and hence on the hardening. An increase in $\lambda$ increases the extent of the resolved shear coherency stresses in the slip plane of the dislocations (Pachaury et al., 2022a). However, their effects on the stress-strain behavior may not be evident due to the small size of the nanopillars. As a result, the dislocations do not interact with the composition fields significantly before encountering the boundaries of the nanopillars. To further confirm this, we have carried out a separate set of simulations for the inhomogeneous alloy (I) configurations to determine the effects of only the composition fluctuations on the stress-strain behavior of the alloys. This is done to make sure that that any effect coming from the irradiation loops are eliminated while determining the dependence of the stress-strain behavior on λ. Similar to the IL alloy configuration, we did not observe any significant changes in the stress-strain behavior as a function of λ for inhomogeneous (I) alloy configuration. Hence, the analysis presented in this paper for taking the effects of the composition fluctuations can be considered to be representative of the composition fluctuations in irradiated alloy despite artificial composition covariance relaxation length scales used. Lastly, we present the results for the orientation dependent loading response of the irradiated (IL) nanopillars.

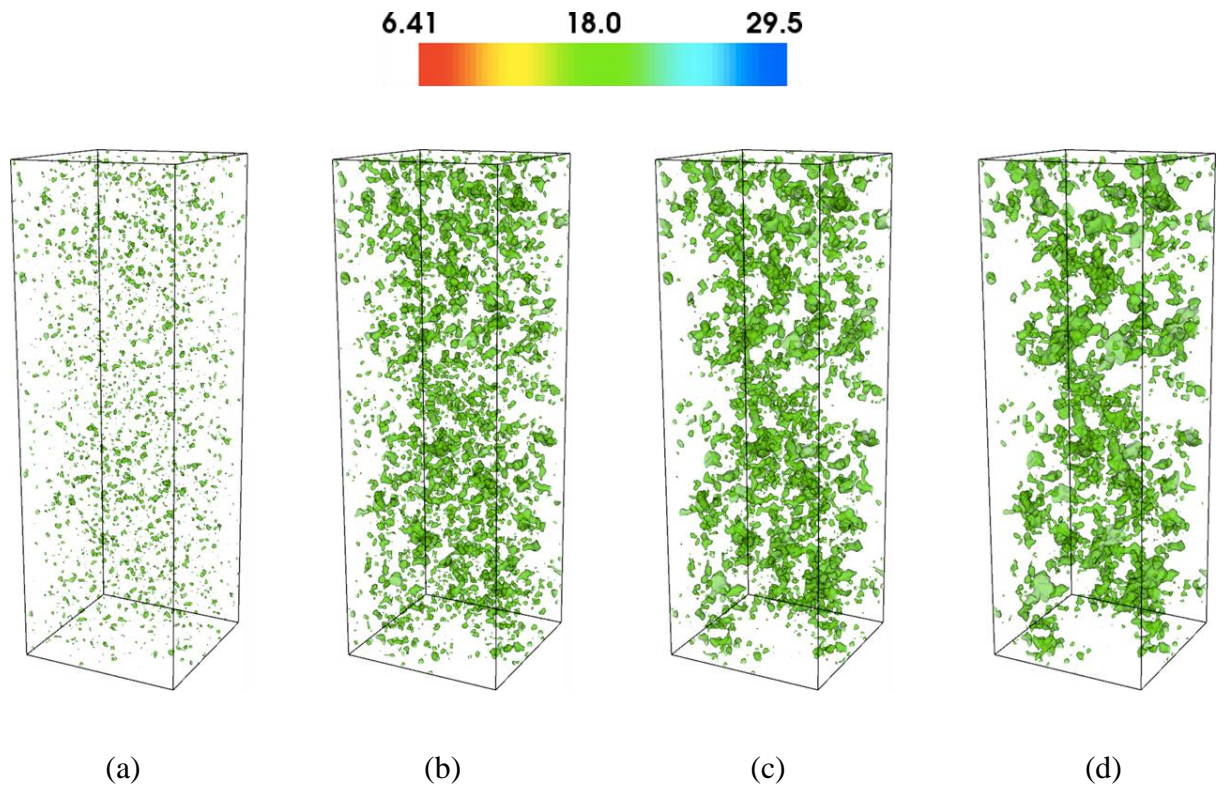

(a)          (b)          (c)          (d)

Figure 10: 18 wt.% iso-surface Cr composition corresponding to (a) 10 nm, (b) 30 nm, (c) 50 nm, and (d) 75 nm composition covariance relaxation length scale (λ).

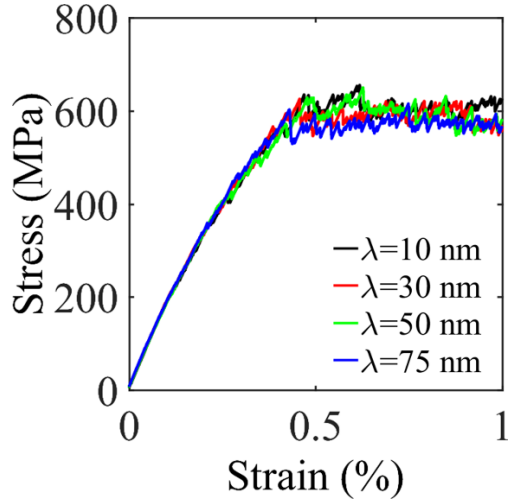

Figure 11: Stress-strain response of the irradiated (IL) nanopillars as a function of composition covariance relaxation length scales (λ).

Figure 12 shows the stress-strain behavior and 0.2% yield strength of irradiated FeCrAl nanopillars (IL) in different loading orientations. Irradiated (IL) nanopillars exhibit different stress-strain responses along different loading orientations indicating varying dislocation activity. [001] loading orientation gives the smallest yield and hardening response followed by [110] loading orientation whereas [111] loading orientation exhibits highest yield strength and hardening among the three orientations tested. Orientation dependent differences in the stress-strain behavior of the irradiated (IL) nanopillars can be explained in terms of the Schmidt factor corresponding to the activated slip-systems (Fang et al., 2022). [001] and [110] loading orientations result in the activation of 8 and 4 slip-systems, respectively, having a Schmidt factor of 0.41, whereas [111] loading orientation results in the activation of 6 slip-systems with a Schmidt factor of 0.27. As compared to [110] orientation, larger dislocation activity is expected in [001] loading orientation. This is expected to be the reason for the observation of smallest yield strength in the [001] loading orientation. Although the number of slip-systems that will be activated is larger in [111] as compared to the [110] orientation, the Schmidt factor is smallest in the [111] loading orientation which implies that more stress will be required to accumulate plastic deformation in the activated slip systems. This is expected to result in the largest yield strength in the [111] oriented crystal. As compared to the [001] loading orientation, similar dislocation activity is observed in [110] and [111] loading orientations where the glide dislocations interact with the irradiation induced defects thereby resulting in formation of pinning points which may lead to generation of single arm sources. Destruction of the irradiation loops due to their interactions with the glide dislocations is also observed. Figure 13 shows one such interaction between an $\frac{a}{2}[\bar{1}\bar{1}\bar{1}]$ type irradiation loop and an $\frac{a}{2}[111]$ type straight glide dislocation which resulted in the generation of a single arm source in the crystal. The glide dislocation interacts with the irradiation loop (dislocations prior to the interaction are shown in Figure 13(a)) which results in partial annihilation of the loop and leads to the formation of pinning points. The large straight dislocation pinned along the irradiation loops result in the formation of a single arm source as shown in Figure 13(b) which emanate dislocations into the crystal as the deformation proceed as shown in Figure 13(b)-(d).

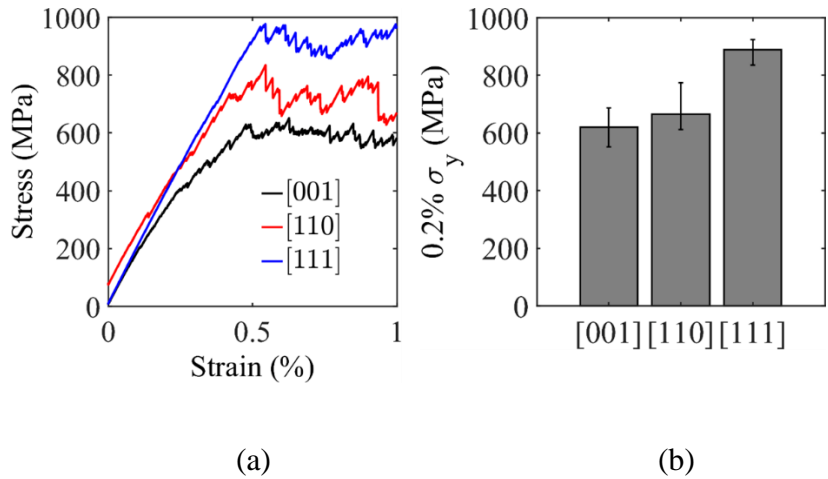

(a)                     (b)

Figure 12: Orientation dependent (a) stress-strain response, and (b) 0.2% yield strength of irradiated (IL) FeCrAl nanopillars

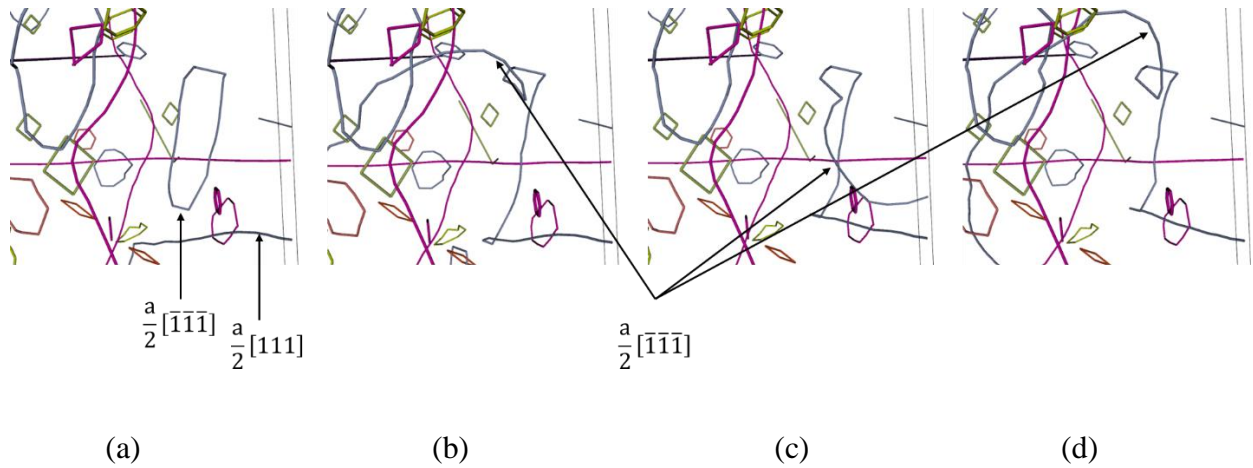

Figure 13: Dislocation activity during [110] loading orientation: (a) Interaction between an $\frac{a}{2}[\bar{1}\bar{1}1]$ type irradiation loop and an $\frac{a}{2}[111]$ type straight glide dislocation, (b)-(d) the interaction leads to the partial destruction of the irradiation loop due to its partial annihilation with the glide dislocation. This results in the formation of a single arm source of type $\frac{a}{2}[\bar{1}\bar{1}\bar{1}]$. Activity of the single arm source with time is also shown (b)-(d).

## 4. Conclusions

In this paper, DDD has been utilized to investigate plasticity in irradiated FeCrAl nanopillars. This paper is a first attempt towards characterizing the effects of different irradiation-induced defects separately as well as superposed together using mesoscale simulations. Irradiation-induced dislocation loops and composition inhomogeneity are taken into consideration for a holistic representation of irradiated FeCrAl microstructures. Irradiation-induced loops are modeled as closed-line dislocations whereas composition fluctuations are modelled using a stochastic process by utilizing 2D composition fields acquired from EDX data of irradiated FeCrAl alloys. Subsequently, composition dependent mobility and internal stress fields, determined from finite element solution of an elastic boundary value problem, are integrated into DDD. It is observed that irradiated alloys exhibit higher yield strength as compared to the homogeneous alloys. Moreover, a "destructive interference" is observed in the hardening contributions of different irradiation induced defects wherein the composition inhomogeneity superposes in an anti-correlative manner with the irradiation induced loops. This happens due to coherency stresses changing the morphology and orientations of the $a/2\langle 111\rangle$ type irradiation loops. This "destructive interference" in the hardening contributions due to different irradiation induced defects has not been reported before. This anti-correlative superposition may be able to explain our parallel TEM *in situ* tensile tests on unirradiated, ion irradiated, and neutron irradiated C35M FeCrAl alloy and may aid the experimental observations to better determine the hardening contributions from analytical hardening models. Hardening in nanopillars is determined to be due to dislocation starvation and subsequent activation of dislocation sources at higher stresses. Irradiation induced defects act as barriers to the motion of dislocations. These defects change the topology of the dislocation network through formation of pinning points, dislocation sources, and junctions, the effect of which is manifested in the mechanical behavior of the irradiated alloys. Effects of the loading orientations on the yield strength and hardening are also investigated where [111] loading orientation exhibits the highest yield strength followed by [110] and [001] loading orientations. The methodology presented in this paper will serve as a benchmarking tool for assessing properties of FeCrAl alloys targeted towards nuclear energy applications.

**Acknowledgements** – This work was supported by the U.S. Department of Energy, Office of Nuclear Energy (DOE-NE) through Nuclear Energy University Programs contract DE-NE0008758.